\documentclass[aps,pre,twocolumn,floatfix]{revtex4}
\usepackage{graphicx}
\usepackage{amssymb,amsmath,amsthm}
\usepackage[all]{xy}
\newcommand{\rt}{\rightarrow}
\def\be{\begin{equation}}
\def\ee{\end{equation}}
\def\bea{\begin{eqnarray}}
\def\eea{\end{eqnarray}}

\def\taj{\tilde a_{j}}
\def\ta0{\tilde a_0}
\def\ta{\tilde a}

\begin{document}

\title{Intrinsic noise induced resonance in presence of sub-threshold signal in Brusselator}

\author{Supravat Dey}


\affiliation{Department of Physics, Indian Institute of Technology Bombay, Powai, Mumbai-400076, India}

\author{Dibyendu Das}


\affiliation{Department of Physics, Indian Institute of Technology Bombay, Powai, Mumbai-400076, India}

\author{P. Parmananda}


\affiliation{Department of Physics, Indian Institute of Technology Bombay, Powai, Mumbai-400076, India}


\begin{abstract} 
{In a system of non-linear chemical reactions called the Brusselator, we show that {\it intrinsic noise} can be regulated to drive it to exhibit resonance in the presence of a sub-threshold signal. The phenomena of periodic stochastic resonance and aperiodic stochastic resonance, hitherto studied mostly with extrinsic noise, is demonstrated here to occur with inherent systemic noise using exact stochastic simulation algorithm due to Gillespie. The role of intrinsic noise in a couple of other phenomena is also discussed.} 
\end{abstract}



\maketitle

\noindent{\bf Studies of noise driven regularity in non-linear systems have 
not been done as extensively for ``intrinsic noise" as for extrinsic noise. 
Here we give direct demonstration of stochastic resonance (both periodic and 
aperiodic) in a chemical system with respect to intrinsic systemic 
fluctuations, using exact stochastic simulation method due to Gillespie.
Moreover, an interplay of the intrinsic  and extrinsic noises is
analyzed for these noise invoked resonances.}

\section{Introduction}
Chemical and biological systems quite generally are stochastic in nature. The stochasticity may arise from  the inherent probabilistic nature of the processes involved or external environmental interferences --- accordingly they are referred to as ``intrinsic" and ``extrinsic" noise, respectively.  For a chemical system specified by certain reaction rate constants, the chemical reactions in actual reality vary randomly, and molecular numbers do fluctuate --- this is intrinsic to the system \cite{delbruck,bartholomay}. Similarly biological processes like gene expression \cite{mukund,swain} involve intrinsic stochasticity. As opposed to this, if fluctuations in the medium or other chemical components external to the subsystem of interest, indirectly affect the biochemical processes involved, they are regarded as extrinsic. Although for biochemical systems with large number of molecules, deterministic approximations of chemical reactions are often enough, for systems with small number of molecules, intrinsic fluctuations are relatively large and stochastic treatment is necessary. In the latter case, the system's response may go beyond mere small excursions about the averages and unexpected behavior may appear due to random crossing of thresholds. A theoretical framework within which intrinsic noise in biochemical systems is studied is the chemical master equation (CME) approach. A widely used numerical method to exactly implement the CME is Gillespie algorithm \cite{gillespie76,gillespie77,gillespieRev}, as the CME is often not easily analytically tractable.     

Noise, intrinsic or extrinsic, is commonly thought of as an undesirable disturbance. Yet in the realm of non-linear systems, it is well known now that noise can aid the output attain regularity, or at times, reveal hidden order. Here by ``regularity" we imply either enhanced periodicity of the output response, or enhanced correlation of the output response to the input signal. In this paper, we study primarily periodic stochastic resonance (PSR) and aperiodic stochastic resonance (ASR). These effects have been seen in multiple non-linear systems \cite{Benzi,Benzi2,Nicolis2,collins1,collins2,eichwald,Gammaitoni,punitprl04,punitasr} with extrinsic noise. When a non-linear system is near a bifurcation threshold, and subjected to an weak input signal superposed with a noise, the response of the output shows maximum ``regularity" for an optimum strength of noise. The input signal used for PSR is periodic and for ASR is aperiodic. The superimposed noise used in earlier studies were mostly extrinsic --- for example, in the theoretical models \cite{Benzi,Benzi2,Nicolis2} noise has been added externally to the dynamical equations, while in experiments on electrochemical cell \cite{punitprl04,punitasr} external noise has been added to the voltage. 

A natural question is that can these phenomena be seen as a result of variation of {\it intrinsic noise} inherent to the system, near a bifurcation threshold? Experimentalists vary different sources of internal noise to study regularity of response, suitable and specific to their system of interest --- e.g. in an electrochemical cell chloride ion concentration was varied \cite{punitintnoise}, while in hippocampal CA1 neurons sub-threshold cathodic current was used \cite{william}. Although sources of internal noise can be diverse in practical systems, for keeping our theoretical discussion general, we follow the standard ideas of intrinsic stochasticity in chemical reactions used by Gillespie \cite{gillespie76,gillespie77}. The idea is to simulate CME numerically using {\it exact} stochastic simulation algorithm given by Gillespie \cite{gillespie76,gillespie77}. Instead of the latter direct approach, in the literature, often chemical Langevin equation (CLE) \cite{gillespieCLE, gillespieRev} has been used \cite{xin,xin1,hanggi,Anandi,hanggi}. It is important to note that CLE is an approximation of CME \cite{cleassumption}, and sometimes produce misleading results \cite{baras}. Therefore, CME is preferable. In this paper we implement CME using the Gillespie method.

The phenomenon of coherence resonance (CR) \cite{pikovsky,giacomelli,Lindner,Neiman,santiPre}, where there is no input sub-threshold signal, has been studied earlier using Gillespie and CLE method \cite{xin,xin1,xin3,xin4,xin5,xin6,hanggi}. To our knowledge, demonstration of PSR and ASR in a chemical system with Hopf bifurcation using exact CME approach is not well known; although this has been studied for bistable nonlinear chemical systems \cite{leonard,dykman}. In the present work, we study PSR and ASR in the chemical Brusselator system using Gillespie algorithm. Furthermore, we explore the effect on PSR and ASR due to the interplay of extrinsic and intrinsic noises. Apart from that, we revisit the CR and another noise driven phenomenon to make some interesting observations.  
            
The Brusselator oscillator \cite{Prigogine,gillespie77,Anandi} system involves the following four chemical reactions.
\bea
\nonumber
Z_1 &\stackrel{k_1}{\rt}& X~~~~:Z_1~{\rm fixed}\\
\nonumber
Z_2 + X &\stackrel{k_2}{\rt}&  Y + Z_3~~~~:Z_2~{\rm fixed}\ \\
\nonumber
2 X + Y &\stackrel{k_3}{\rt}&  3 X \\
X &\stackrel{k_4}{\rt}& Z_4 
\label{bruss}
\eea
$Z_1$ and $Z_2$ are kept fixed by having infinite sources of these reactants. In Eq. \ref{bruss}, \{$k_j$\} ($j$ = 1,2,3,4) are the deterministic  reaction rate constants for the four reactions respectively. The symbols $X,~Y,~Z_1,~Z_2,~Z_3,~Z_4$ represent the molecular numbers for the six chemical species involved. The system volume is denoted by $\Omega$. 
The numbers of $X,~Y,~Z_3$ and $Z_4$ vary with time; out of them $X$ and $Y$ are the primary variables of interest. For certain choice of the reaction rate parameters, the reactants  $X$ and $Y$  fluctuate around fixed point values, while other choice of rates, they fluctuate around limit cycle behavior. These two regimes are separated by a Hopf bifurcation. Within Gillespie algorithm, propensity functions \cite{gillespie76,gillespie77} for the above reactions are,
\bea
\nonumber
\ta_1=Z_1k_1,~
\ta_2=Z_2Xk_2/\Omega,\\
\ta_3=X(X-1)Yk_3/\Omega^2,~{\rm and}~
\ta_4=Xk_4.
\label{propensity} 
\eea
respectively. 

We discuss the Gillespie algorithm \cite{gillespie76,gillespie77,gillespieRev} for this system briefly. As $Z_1$ and $Z_2$, coming from infinite sources, are not time varying, we consider a state vector ${\bf s}(t)$ excluding them, namely ${\bf s}(t)\equiv (X(t),~Y(t),~Z_3(t),~Z_4(t))$. Starting with a state vector ${\bf s}(t)$ at time $t$, the $j^{th}$ reaction out of the four in Eq. \ref{propensity} is chosen with a probability $\ta_j/\ta_0$, after a waiting time $\delta t$, drawn from a probability distribution function $P(\delta t | {\bf s}(t))=\ta_0\exp(-\ta_0\delta t)$. Here \{$\ta_j$\} is given by Eq. \ref{propensity}, and $\ta_0=\sum_{j=1}^{4}\taj$, and these are functions of ${\bf s}(t)$. After the implementing $j^{th}$ reaction, the state vector ${\bf s}(t)$ is updated to ${\bf s}(t+\delta t)=${\bf s}(t)$+{\bf\nu}_j$, where using Eq. \ref{bruss} one can easily see that the state change vectors, ${\bf\nu}_1=(1,0,0,0)$, ${\bf\nu}_2=(-1,1,1,0)$, ${\bf\nu}_3=(1,-1,0,0)$, and ${\bf\nu}_4=(-1,0,0,1)$.

\begin{center}
\begin{figure}[]
\includegraphics[scale=0.83]{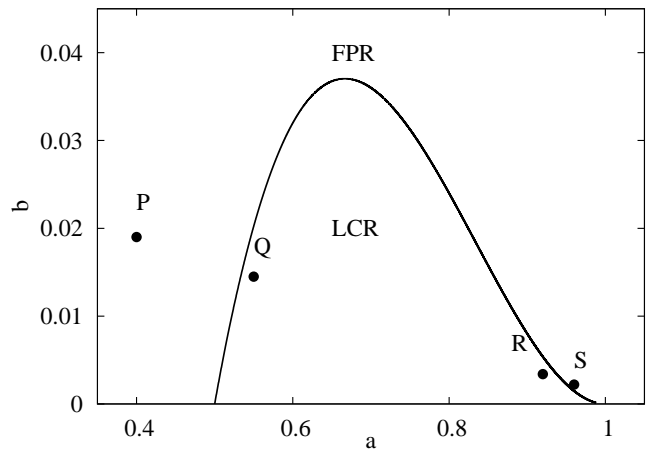}
\caption{\label{ab}Shows the limit cycle regime (LCR) and the fixed point regime (FPR) in `a-b' parameter space for Brusselator. Dynamical behavior of the points  P, Q, R, and, S in the deterministic and the stochastic cases are shown in Fig. \ref{path}. }
\end{figure}
\end{center}

The way to tune intrinsic noise strength to study resonances for this system, is to vary $\Omega$, $Z_1$ and $Z_2$, keeping the concentrations $z_1=Z_1/\Omega$ and $z_2=Z_2/\Omega$ fixed by hand. Under the latter condition, we observed that the numbers {$X,~Y$} also spontaneously adjust to keep their concentrations $x=X/\Omega$ and $y=Y/\Omega$, on an average close to a constant. The observation is expected since the concentration $x$ and $y$, if approximated to be deterministic, satisfy the following equations: \bea
\nonumber
\frac{dx}{dt} &=& k_1 z_1 - k_2 z_2 x + k_3 x^2 y - k_4 x\\
\frac{dy}{dt} &=& k_2 z_2 x - k_3 x^2 y.
\label{bruss1}
\eea
The above equations show that if the concentration $z_1$ and $z_2$ are fixed, the concentration $x$ and $y$ can be determined. Thus by increasing (decreasing) $\Omega$, $Z_1$ and $Z_2$, keeping concentrations $z_1$ and $z_2$ fixed, we ensure that we stay at the same parameter set point of the non-linear system, and keep decreasing (increasing) the ``intrinsic noise" around it. The condition for getting limit cycle behavior for the deterministic Eq. \ref{bruss1} is \cite{Anandi},
\be
b~<~(2a-1)(1-a)^2
\label{abrelation}
\ee
where $a = k_2z_2/(k_2z_2 + k_4)$ and $b = (k^2_1z^2_1k_3)/[(k_2z_2+k_4)^3)]$. In Fig. \ref{ab}, the solid line is Hopf bifurcation line separating fixed point regime (FPR) from the limit cycle regime (LCR). 

Although our actual simulation is stochastic following Eq. \ref{bruss}, in order to identify the average position of the set point of the system, we fall back whenever necessary to the Eq. \ref{bruss1}. By this we mean that, a particular choice of $k_1$, $k_2$, $k_3$,  $k_4$,  $z_1$ and $z_2$, corresponds to a particular $a$ and $b$ value, and the Fig. \ref{ab} tells us the average location of the set point of the system. Thus using Fig. \ref{ab} as guidance, we proceed to check the phenomena of PSR and ASR.

\section{Results}

\begin{center}
\begin{figure*}[!htbp]
\includegraphics[height=3.5in,width=7in]{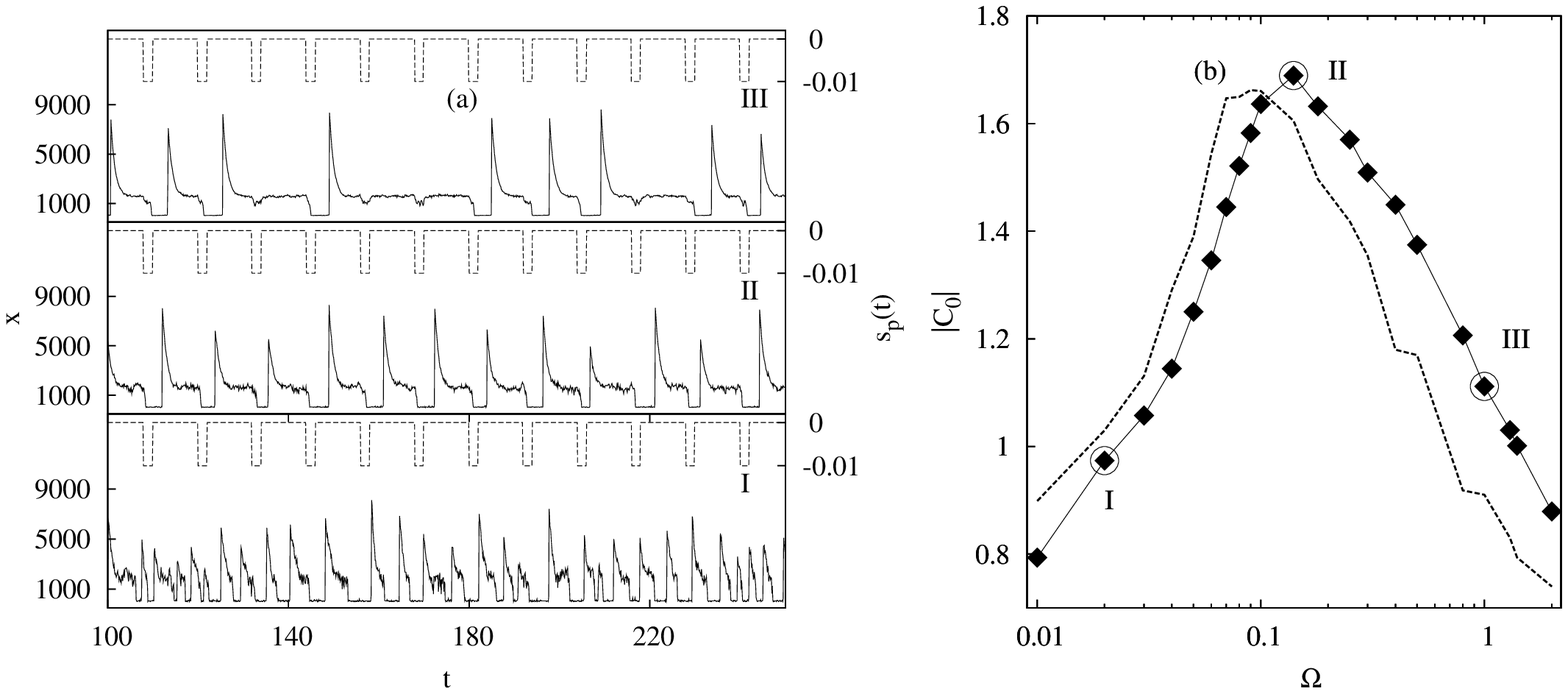}
\caption{Data for PSR. (a): The lower curves in each box show behavior of $x(t)$ for increasing system volumes (decreasing noise) : $\Omega = 0.02$ (I),  $\Omega = 0.14$ (II), and $\Omega = 1.0$ (III), in the presence of sub-threshold periodic signal $s_p(t)$ (the upper curves in every box). Here the pulse amplitude is -0.010, the interpulse interval is 10.0, and pulse width is 2.0. We observe that output spikes become more correlated  with the input periodic signal in case II, compared to I and III. Note that an output spike begins with a reduction from the steady value, followed by a rapid rise --- thus there is no significant phase shift between an input pulse and the output spike, as may appear at a first glance. Panel (b): The solid line with points shows absolute value of power norm $C_0$ against $\Omega$ (pure intrinsic noise); the dashed line shows the same, for an additional extrinsic noise (see the text). Note that $|C_0|$ has a peak at a smaller value of $\Omega$ for the added extrinsic noise. \label{psr}}
\end{figure*}
\end{center}

\begin{figure*}[!htbp]
\includegraphics[height=3.5in,width=7in]{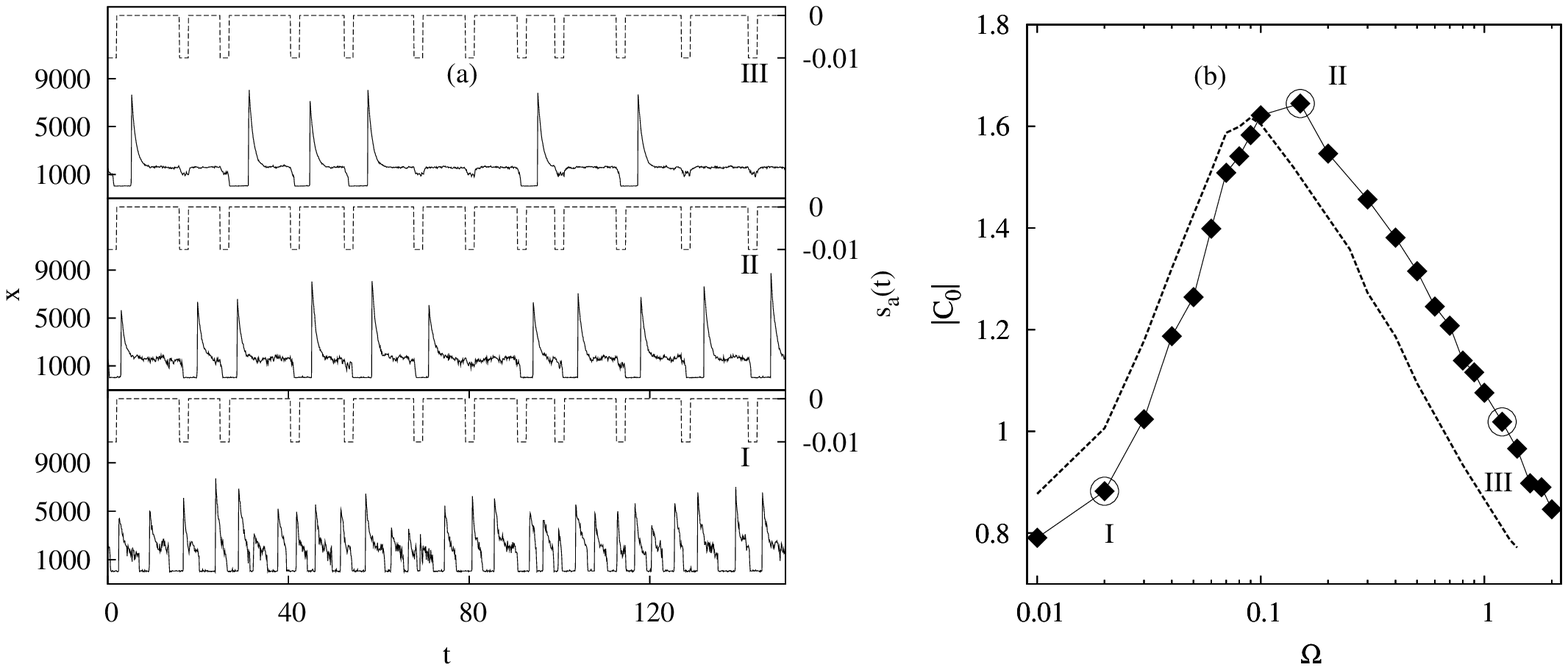}
\caption{Data for ASR. Panel (a): The lower curves in each box show behavior of $x(t)$ for increasing system volumes (decreasing noise) : $\Omega = 0.02$ (I),  $\Omega = 0.15$ (II), and $\Omega = 1.40$ (III), in the presence of sub-threshold aperiodic signal $s_a(t)$ (the upper curves in every box). Here the pulse amplitude is -0.010, the pulse width is 2.0. Interpulse intervals are aperiodic and its exact description is given in the text. We observe that output spikes become more correlated with the input aperiodic signal in case II, compared to I and III. As commented in Fig. \ref{psr}, there is no significant phase shift between an input pulse and the output spike, as may appear at a first glance. Panel (b): The solid line with points shows absolute value of power norm $C_0$ against $\Omega$ (pure intrinsic noise); the dashed line shows the same, for an additional extrinsic noise (see the text). Note that in this case also $|C_0|$ has a peak at a smaller value of $\Omega$ for the added extrinsic noise.
 \label{asr}}
\end{figure*}

In this section we show that the phenomena PSR and ASR happen due to intrinsic noise of the Brusselator system. In each case our set point is in the FPR,  represented schematically by S shown in Fig. \ref{ab}. We tune the intrinsic noise strength for these reactions by varying $\Omega$, $Z_1$ and $Z_2$, such that the concentrations $z_1=Z_1/\Omega$ and $z_2=Z_2/\Omega$ are held fixed by hand as discussed above. The noise makes the system occasionally excurse into LCR from FPR and one sees spikes in $x$ and $y$ as a result of that. For different noise strengths we have looked at the output response  $x$ and $y$, but in the subsequent discussion  we will focus only on the variable $x$ for brevity.

To keep the discussion simple we will suppress the explicit dimensions of the rate constants $\{k_j\}$, and the volume $\Omega$. This makes us avoid putting any units for $x(t),~t,~V_N~$, and $C_0$ in all the subsequent figures 2-5. Furthermore, we denote the strength of the intrinsic noise by $\Omega$ --- {\it the larger the $\Omega$, the smaller the intrinsic noise}. Thus in all the figures 2-4, noise decreases along the increasing $\Omega$ axis.  

To characterize the regularity in the response of the system, a measure of the  correlation between output response and input signal, called the power norm $C_0$ \cite{eichwald,punitasr} is used:
\be
C_0 = \langle (x(t) - \langle x\rangle)(s(t) - \langle s\rangle)\rangle.
\label{c0}    
\ee
Here $\langle.\rangle$ denotes the time average. The variable $x(t)$ is the time varying output of the system at time t, and $s(t)$ is the input signal. The signal $s(t)$ is periodic for PSR and aperiodic for ASR. $C_0$ can be positive or negative depending on the relative sign of output response and the input signal. In this study $C_0$ is always positive, but for easier comparison with earlier publications \cite{eichwald,punitasr} we use $|C_0|$.

We will decide the average location of the set point of our system, in `a-b' space by choosing appropriate $k_1,~k_2,~k_3,~k_4,~z_1$, and $z_2$ values. In our simulation of PSR and ASR, discussed below, we take $k_2 = 0.1$, $k_3 = 0.00005$,  $z_1=100$ and $z_2=500$. The values of $k_1$ and $k_4$ are different for different cases and are mentioned in the respective subsections.

\subsection{Periodic Stochastic resonance}
\label{subpsr}

To study the PSR, we choose $k_1$ and $k_4$ to be time dependent: $k_1(t)=k_2z_2(1-a_0-s_p(t))/(a_0+s_p(t))$ and $k_4(t)=[b_0/(z_1^2k_3(a_0+s_p(t)))]^{1/2}$. Here $s_p(t)$ is a time varying {\it periodic} signal. By doing periodic modulation of $k_1$ and $k_4$ with time, the average location $(a,b)$ of the set point gets modulated periodically with time and is given as ($a_0+s_p(t),b_0$) --- see the relations just below Eq. \ref{abrelation}. Thus $s_p(t)$ is to be regarded as a periodic input to the system. We take $a_0=0.980$, $b_0=0.001$, and the amplitude of $s_p$ is -0.010, such that minimum value of $a=0.980-0.010=0.970$, is still greater than the threshold value $a_{th}=0.967$ of the Hopf bifurcation point $(a_{th},b_0)$ in the absence of noise. The sub-threshold signal amplitude value is $77$ percentage of the threshold value (separation between the average set point position and bifurcation point). For these parameters and the input signal we run our Gillespie simulation for different values of $\Omega$, $Z_1$, and $Z_2$, thereby varying the intrinsic noise strength. Fig. \ref{psr}(a) shows output response ($x$ variable) for three $\Omega$ values --- (I) low, (II) intermediate and (III) high. We see that for an intermediate $\Omega$ (II), the output spike pattern becomes more correlated with the input periodic pulse pattern, as compared to the cases I and III --- this is the point of resonance. From the time series $x(t)$ and $s_p(t)$, we calculate $C_0$ using Eq. \ref{c0}. The solid line with points in Fig. \ref{psr}(b) shows $|C_0|$ against $\Omega$, in the absence of any extrinsic noise. An unimodal behavior is seen, where the maximum (implying maximal regularity) corresponds to the optimum value of $\Omega=0.14$. Thus with intrinsic noise we have demonstrated PSR in the Brusselator system.

If extrinsic noise was present in the system, would the internal noise causing the phenomenon of PSR, act to reinforce or subdue the former effect? To explore this interesting question of interplay between extrinsic and intrinsic noises, we proceed to add a noise term to the average set point position: $a_0\rightarrow a_0 + D\xi$. We choose extrinsic noise $\xi$ to be a uniform box distribution between $[0,1]$ and $D = 0.002$. Thus the noise is one sided and pushes (randomly though) the average set point further away from the bifurcation point. The question is whether we will require larger or smaller internal noise to achieve PSR in the presence of finite $D$. The dashed curve in Fig. \ref{psr}(b) shows that the resonance occurs at a smaller value of $\Omega$ (i.e. larger intrinsic noise). Thus we observe that the two sources of noise, extrinsic and intrinsic, actually superimpose.

\subsection{Aperiodic Stochastic resonance}
ASR is similar to PSR, except that the signal $s_a$ is aperiodic. The input aperiodic signal used by us is an aperiodic pulse train. Aperiodicity comes only in the interpulse interval of the signal, which given by $6.0+8.0\times r$. Here r is a random number drawn from uniform box distribution between 0 to 1. Note that the width of the pulses are the same as $s_p$. We take the same parameters as in PSR, except the time varying $k_1(t)=k_2z_2(1-a_0-s_a(t))/(a_0+s_a(t))$ and $k_4(t)=[b_0/(z_1^2k_3(a_0+s_a(t)))]^{1/2}$. The pulse amplitude remains the same as for $s_p(t)$ (see section \ref{subpsr}) such that the average set point location remains greater than $a_{th}=0.967$. Again we vary the intrinsic noise by varying $\Omega$, $Z_1$, and $Z_2$. The Fig. \ref{asr}(a) shows output response ($x$ variable) for three values of $\Omega$: (I) high, (II) intermediate and (III) large. We see that for an intermediate $\Omega$ (II) output spike pattern become more correlated with the input aperiodic pulse pattern as compared to I and III --- this is the point of resonance. From the time series we calculate $C_0$ using Eq. \ref{c0}, and the solid line with points in Fig. \ref{asr}(b) shows $|C_0|$ against $\Omega$, in the absence of extrinsic noise. The maximum regularity corresponds to the optimum $\Omega=0.15$, just like PSR. Thus with intrinsic noise we have demonstrated ASR in the Brusselator system.  

Analogous to PSR, in this case too we explore the interplay between extrinsic and intrinsic noises. We add a noise term to the average set point position: $a_0\rightarrow a_0 + D\xi$. Extrinsic noise $\xi$ is uniformly distributed over $[0,1]$ and $D = 0.002$. The dashed curve in Fig. \ref{asr}(b) shows that the resonance occurs at a smaller value of $\Omega$ (i.e. larger intrinsic noise). Thus for ASR, we also observe that the extrinsic and intrinsic noises superimpose.

\subsection{Test of a mathematical formula for coherence resonance}
\label{seccr}
As noted, coherence resonance (CR) has been studied earlier with intrinsic noise using exact stochastic simulations \cite{xin3,xin4,xin5,xin6}. In this sub-section, we revisit this phenomenon to test a semi-analytical formula for normalised variance $V_N$ recently proposed \cite{santiPre}. In the case of CR there is no input signal, i.e $s(t)=0$, and so $C_0$ (Eq. \ref{c0}) cannot be used as a measure of regularity. Instead normalized variance $V_N$ \cite{pikovsky,santiPre} is used: 
\be
V_{N} = {\sqrt{{\langle \tau_{p}^2 \rangle} - 
{\langle \tau_{p} \rangle}^2}}/{\langle \tau_{p} \rangle}, 
\label{vne1}
\ee
where $\tau_p$ is interspike time intervals and $\langle . \rangle$ again denotes time average. Typically $V_N$ is enumerated from the time-series analysis of spikes generated by the system, using Eq. \ref{vne1}. An alternative formula for $V_N$ as a function of noise strength was proposed recently, which directly reflects the theoretical understanding of the phenomenon of CR. The formula \cite{santiPre} is as follows:
\be
V_{N} = \frac{\tau_{\rm esc}(\Omega)}{\tau_{\rm min}(\Omega) + \tau_{\rm esc}(\Omega)},
\label{vne2}
\ee

 Unlike Eq. \ref{vne1}, this is not a definition of $V_N$, but a suggested mathematical form. It expresses the important fact that CR comes about due to competitive interplay of two time scale $\tau_{\rm esc}$ (a first passage activation time)  and $\tau_{\rm min}$ (an excursion time of virtual limit cycle) \cite{santiPre}. The formula was tested in \cite{santiPre} for the FitzHung-Nagumo system \cite{pikovsky} and a chemical oscillator model \cite{karantonis} in the presence of extrinsic noise. Here we would like to see if it holds good for intrinsic noise. Note again that intrinsic noise strength is proportional to $\Omega^{-1}$, the analog of extrinsic noise strength $D$ \cite{santiPre}. In Eq. \ref{vne2} $\tau_{\rm esc}$ and $\tau_{\rm min}$ can be calculated from the probability distribution function $P(\tau_p)$ of $\tau_p$. The function $P(\tau_p)$ remains zero between $\tau_p=0$ and $\tau_p=\tau_{\rm min}$, and then rises to a peak and eventually decays with an exponential tail for large $\tau_p$ with a time constant $\tau_{esc}$.   
 \begin{center}
\begin{figure}[]
\includegraphics[scale=0.83]{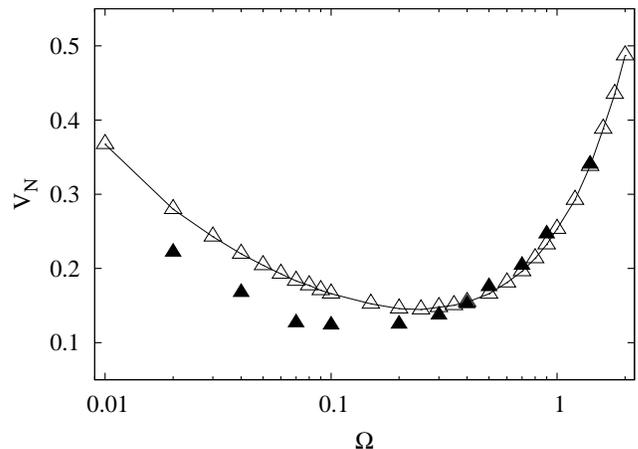}
\caption{\label{vn}Data for CR.  Normalized variance $V_N$ plotted against $\Omega$ --- data from calculation using time series and Eq. \ref{vne1} (symbols $\triangle$ joined by line) alongside data obtained using Eq. \ref{vne2} ($\blacktriangle$). $V_N$ has a minimum for an optimum value of $\Omega=0.25$.}
\end{figure}
\end{center}
 
\begin{center}
\begin{figure*}[htbp]
\includegraphics[height=3.5in,width=7.0in]{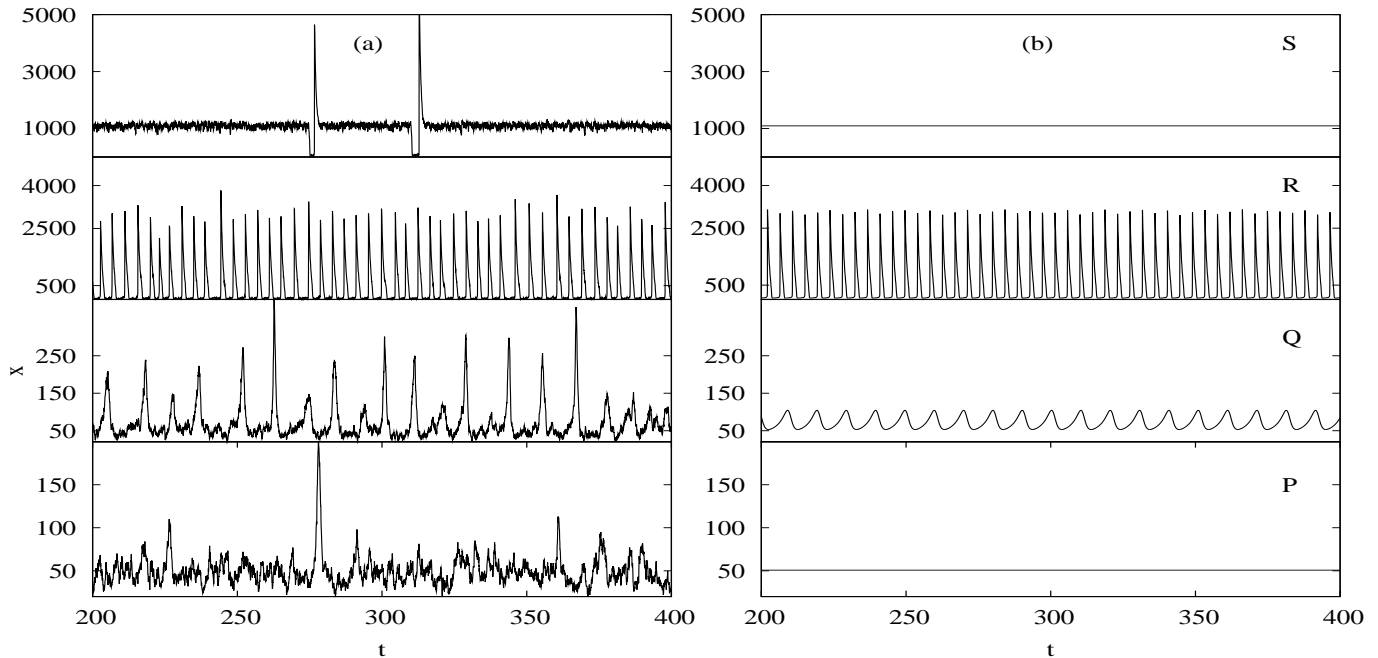}
\caption{Shows behavior of $x(t)$ in four different points P, Q, R, and S (from bottom to top) of `a-b' space. Panel (a) are the output responses in the presence of intrinsic noise and panel (b) are the corresponding deterministic responses. For these data, we used $k_1=0.10$, $k_2=0.10$, $k_3=0.00005$, $k_4=1.75$, and $\Omega=1.0$. The various values of ($Z_1,~Z_2$) pairs used are (888, 11) for P, (1270, 21) for Q, (8421, 201) for R, (19126, 419) for S. \label{path}}
\end{figure*}
\end{center}

We choose $k_1 = 16.551$ and $k_4 = 1.546$ which correspond to $a=0.970$ and $b=0.001$, setting the average location of the set point of our system in the FPR. $V_N$ was calculated from the time series analysis of the output signal $x(t)$ for various $\Omega$ --- the curve is plotted in Fig. \ref{vn} (with $\triangle$ symbols joined by a continuous line). Next, we calculated $P(\tau_p)$ and then found $\tau_{\rm esc}$ and $\tau_{\rm min}$ from it ---  we used the latter to calculate  $V_N$ using Eq. \ref{vne2} (shown with $\blacktriangle$ symbols in Fig. \ref{vn}). The curves of $V_N$ obtained by two different procedures match reasonably well confirming the validity of Eq. \ref{vne2}, in the case of intrinsic noise in Brusselator.

\subsection{Output response on varying reactant molecular numbers, keeping $\Omega$ fixed.}
\label{comment}
In certain situations, it may be inconvenient to keep reactant concentrations fixed. It may be convenient to vary just the reactant numbers, or just the volume, independently. Does any interesting phenomenon arise in such cases? Here we consider varying numbers $Z_1$ and $Z_2$, keeping $\Omega$ fixed. This makes average concentrations of the reactants $z_1$, $z_2$, $x$, and $y$ vary, and hence make the average set point of the system drift. The drift carries the system from FPR through LCR, and the noisy observable $x(t)$ shows interesting change in temporal pattern due to that. The latter evolution is explicitly shown in Fig. \ref{path}(a). By tuning $Z_1$ from 888 through 19126, and $Z_2$ from 11 through 419 (with $\Omega=1.0$), the $x(t)$ behavior changes from the bottom to the top of Fig. \ref{path}(a). The spiking behavior of $x(t)$ passes successively from irregular$\rightarrow$ regular $\rightarrow$ irregular, reminding us of the phenomenon of CR. But it should be noted that this phenomenon has nothing to do with CR. The behavior seen in Fig. \ref{path}(a) are noisy excursions on top of the steady behavior at P, Q, R, and, S (see Fig. \ref{ab}), respectively. Due to the change of concentrations, the system drift from FPR$\rightarrow$ LCR$ \rightarrow$ FPR. To substantiate the latter assertion, we show the deterministic steady behavior at P, Q, R, and S in Fig. \ref{path}(b) corresponding to their counterparts in Fig. \ref{path}(a) --- the similarity is obvious. Thus the above way of varying reactant concentrations  keeping $\Omega$ fixed, not only varies internal noise, but also make the set point of the system drift along with it. The effect is interesting but this procedure is unsuitable for tuning inherent noise to study resonance phenomena in the system.

\section{Discussion}
In this paper, we have shown that an optimal value of the intrinsic noise present in a chemical Brusselator system can induce enhanced ``regularity" in response, seen in phenomena like PSR and ASR, just like extrinsic noise. The intrinsic noise strength has been varied for given set $\{k_j\}$, by varying system volume $\Omega$ and reactants numbers, keeping the concentration of the reactants $z_1~\rm{and}~z_2$ fixed; the average concentrations of $x$ and $y$ spontaneously adjusted to stay constant. In section \ref{comment}, we discussed that if we vary the number of the reactants keeping $\Omega$ fixed, their concentrations vary, leading to drift of the average set point of the system. Interesting variation of temporal pattern may follow due to drift from FPR$\rightarrow$ LCR$\rightarrow$ FPR, but that has nothing to do with the phenomenon of CR. In section \ref{seccr} a simple formula for the normalised variance $V_N$ as function of $\Omega$ (inverse strength of intrinsic noise) was tested.


\begin{thebibliography}{10}
\bibitem{cleassumption}
The approximation involves assumptions that (i) the expected number of
  occurrences of each reaction within the update time interval of the Langevin
  equation is very large, and (ii) the reaction rates remain constant within
  these update time intervals. Assumptions (i) and (ii) are hard to satisfy
  simultaneously when the molecular number of species are small.
\end{thebibliography}
\end{document}